\documentstyle[aps,multicol]{revtex}

\begin{document}
\draft

\def\del{\partial}

\title{QHE of Bilayer Systems in the Presence of Tunneling\\
 --  $\nu=1/q$  case --}
 
\author{Kentaro {\sc Nomura} and Daijiro {\sc Yoshioka}}

\address{ Department of Basic Science, University of Tokyo, 
3-8-1 Komaba, Tokyo,  153-8902, Japan}

\date{\today}
\maketitle

\begin{abstract}
Transport properties of bilayer quantum Hall systems at
$\nu=1/q$, where $q$ is an odd integer, are
investigated.
The edge theory is used for the investigation, since
tunneling  between the two layers
is assumed to  occur on the edge of the sample
because of the bulk incompressibility.
 It is shown that in the case of the independent Laughlin state 
tunneling is irrelevant when $\nu<1/2$ in the low temperature
and long wave length limit.
The temperature dependence of two-terminal conductance
of the system in which only one of the two layers is contacted with
electrode is discussed.
\end{abstract}



\begin{multicols}{2} 
\narrowtext
The fractional quantum Hall(FQH) effect appears in two-dimensional electron
systems in a strong magnetic field.
   Recently a large number of studies have been made on bilayer quantum
Hall states. These systems are two dimensional electron
 systems
in double-quantum-well, and various ground states under
 interlayer interactions have been studied.\cite{smg}
   One of the direction of developments is the transport phenomena.
It has become possible to fabricate bilayer systems in which the two
layers can be contacted separately, and some transport properties
have been studied experimentally.\cite{eis,ohno}
In one experiment, Ohno et al.
 studied transport
 for a systems in which contact was made to one layer while the other was left
 floating.\cite{ohno} When the system was tuned to equal layer densities to
 maximize
 coupling due to tunneling, they found that the transport coefficient were
  usually  nearly
identical to those measured when both layers were connected.
However, near the
integer
quantum Hall plateaux, the Hall resistivity almost doubled in the floating
 layer
case.
Yoshioka and MacDonald studied the influence of interlayer
 tunneling upon the longitudinal and transverse resistance in the framework of
 non-interacting electron and successfully explained
 the behavior at the integer
 plateaux. \cite{dyam,dy}
This experiment is done at 1.5K which is too high to
observe the fractional quantum Hall plateaux.
Since then no one has done experiment in a situation where fractional quantum
Hall plateau is observed.
It is not obvious that the FQH plateaux behave similarly to the IQH plateaux
in the presence of tunneling.
If the behavior is different, it will reflect the properties of the
FQHE edge states.
Therefore it is worth while to study the behavior of the FQH plateau
both experimentally and theoretically.
The purpose of the present paper is to theoretically investigate
what happens to the FQH plateau in the presence of tunneling between
the layers.
We will show that the plateau at $\nu=1/q$, where $q>1$, can behave differently
from that at $\nu=1$.

In this paper we consider bilayer systems at
filling factor $1/q$ each, where $q$ is an odd integer.
The electrodes are attached to only one of the layers, and
influence of the other layer on the two-terminal and Hall conductances
 are investigated.
 The edge states, which determine the transport properties,
are influenced by the coupling, since the excitation at the edge is
gapless.\cite{wen2,wen3}
So we first explain how the bilayer edge is described by the two component
chiral Tomonaga-Luttinger liquid theory in the absence of tunneling.
Then we add the effect of tunneling, which occurs only at the edges.
  The edge state of the multicomponent quantum Hall systems is studied by
McDonald and Haldane.\cite{mh} According to them,
two distinct phase should be considered. They are parameterized by 
 $K$-matrix:\cite{wz}
\begin{equation}
  K = \left(\begin{array}{cc}m\pm 1&m \\ 
                 m&m\pm 1
          \end{array}\right),\quad
  q= \left(\begin{array}{c}1 \\1 \end{array}\right),
\end{equation} and
\begin{equation}
  K = \left( \begin{array}{cc} 
                 2m\pm1 & 0 \\
                 0 & 2m\pm1
                     \end{array} \right),\quad
  q = \left(\begin{array}{c}1 \\1 \end{array}\right),
\end{equation}
where $m$ is even integer which is related to the bulk filling factors of 
the both layers $\nu$ as $\nu = 1/(2m\pm1) = 1/q$. When $\nu=1$ these two
states are equivalent.
 They describe standard hierarchical state and independent Laughlin state,
 respectively. 
 We describe the edge state of bilayer FQH systems as follows:
\begin{eqnarray}
S_0 &=& \int {\rm d}\tau {\rm d}x \ {\cal L}_0[\phi],   \nonumber  \\
{\cal L}_0 &=& \sum_{IJ} \frac{{\rm i}}{4\pi} K_{IJ}
  \frac{\partial \phi_I^+}{\partial \tau}
  \frac{\partial \phi_J^-}{\partial x}          \nonumber     \\
& &\ \ \ \ \ \    + \frac{1}{8\pi}v_{IJ}( \frac{\partial \phi_I^+}{\partial x}
\frac{\partial \phi_J^+}{\partial x} + \frac{\partial \phi_I^-}{\partial x}  
\frac{\partial \phi_J^-}{\partial x} ), 
\end{eqnarray}
where $x$ is the coordinate along the edge, pseudo spin
$I,J= {\uparrow}{\downarrow}$ corresponds to two layers,
$\phi^\pm_I = \phi^u_I \pm \phi^l_I$ with $\phi^u_I(\phi^l_I)$ being
 the edge boson fields propagating along
the upper(lower)boundary of the $I$ layer,
$v_{IJ}$ encodes the non universal interactions, and we assume 
$v_{11}=v_{22},v_{12}=v_{21} $.
%
%

In the charge-pseudo spin basis,\cite{kfp}
$\phi_{c, s}=( \phi_{\uparrow} \pm \phi_{\downarrow})/2 $,
the two-components can be separated:
\begin{eqnarray}
{\cal L}_0   = (\  \frac{{\rm i}}{2{\pi} K_{\rho}}\frac{\partial \phi_c^+}
{\partial \tau}
\frac{\partial \phi_c^-}{\partial x} + \frac{v_c}{4\pi K_{\rho}}
[(\frac{\partial \phi_c^+}{\partial x})^2 
+ (\frac{\partial \phi_c^-}{\partial x})^2]   \nonumber \\ 
  + \frac{{\rm i}}{2{\pi} K_{\sigma}}\frac{\partial \phi_s^+}{\partial \tau}
\frac{\partial \phi_s^-}{\partial x} + \frac{v_s}{4\pi K_{\sigma}}
[(\frac{\partial \phi_s^+}{\partial x})^2 
+ (\frac{\partial \phi_s^-}{\partial x})^2]\  ),    \nonumber \\
\end{eqnarray}
where $K_\rho =\pm 1/(2m\pm1)=\pm\nu, K_\sigma = 1 $ for 
standard hierarchical state, 
 $K_\rho= K_\sigma =\nu$ for independent Laughlin  state, and 
$v_{c,s}=2(v_{11} \pm v_{12})$ are
velocities of charge mode, and spin mode, respectively.\cite{note1}

     On the boundary
electron field operator is given by \cite{wen1,in2}
\begin{eqnarray}
   \psi_I^{u} = \sqrt{\rho_0} \ 
\exp({{\rm i} q \phi_c^{u} \pm {\rm i} \phi_s^{u}}), \nonumber \\
   \psi_I^{l}  = \sqrt{\rho_0} \
 \exp({-{\rm i} q \phi_c^{l} \mp {\rm i} \phi_s^{l}}),
\end{eqnarray}
for standard hierarchical state, and
\begin{eqnarray}
   \psi_I^{u} = \sqrt{\rho_0} \ \exp({{\rm i} \phi_I^{u}/\nu }),\nonumber \\
   \psi_I^{l} = \sqrt{\rho_0} \ \exp({ -{\rm i} \phi_I^{l}/\nu }),
\end{eqnarray}     
for independent Laughlin state.
In these equations $\rho_0$ is electron density in the bulk.

Now
we go on to discuss tunneling between the two layers.
The tunneling is described by the following Hamiltonian.
\begin{equation}
 H_{{\rm tun}} = \int {\rm d}x 
\left(  t(x) \psi_{\uparrow}^{\dagger}(x) \psi_{\downarrow}(x)
   + h.c.  \right),
\end{equation}
where $t(x)$ is the tunneling amplitude.
In the ideal system the tunneling occurs uniformly, and $t(x)$ is independent
of $x$.
On the other hand, if the impurity assisted tunneling is dominant,
$t(x)$ depends on spatial distribution of the impurities.
We first consider the former case.
Then Eq.(7) is written in terms of boson fields as
\begin{eqnarray}
 S_{tun} &=&  \int {\rm d}\tau {\rm d}x \ t\ ( \psi_{\uparrow}^{u {\dagger}}
  \psi_{\downarrow}^u               
     +\psi_{\uparrow}^{l {\dagger}} \psi_{\downarrow}^l + \ h.c )
                                    \nonumber \\
     &=&    u \int_{}^{} {\rm d}\tau 
{\rm d}x \ \cos(\frac{\phi_s^+}{K_{\sigma}}  ) \
                                \cos(\frac{\phi_s^-}{K_{\sigma}} ),
\end{eqnarray}
  where $u= 4\rho_0 t$
is coupling constant.

 We suppose the limit where the tunneling amplitude is very weak,
and calculate the scaling dimension of the tunneling amplitude.
 We introduce a energy-momentum cut off $\Lambda$, then bosonic column vector field is
given as
\begin{equation}
    \phi_{\Lambda}({\sf x}) = \int_{0 <|{\sf p}|<{\Lambda} } ^{}
      \frac{{\rm d}^2 {\sf p}}{(2\pi)^2} \ 
{\rm e}^{{\rm i}{\sf p}\cdot{\sf x}} \ \phi ({\sf p})
,
\end{equation}
where ${\sf x}=(\tau,x) $ is $1+1$ dimensional space-time vector,
and ${\sf p}=(\omega,k) $ is energy-momentum vector.
We divide the field $\phi_\Lambda$ into slow and fast modes such
as
\begin{equation}
   \phi_{\Lambda} ({\sf x}) = \phi_{\Lambda'} ({\sf x}) + h({\sf x}),
\end{equation}
 where
\begin{equation}
h({\sf x}) = \int_{ {\Lambda'} <|{\sf p}|< {\Lambda} }^{}
 \frac{{\rm d}^2{\sf p}}{(2\pi)^2}
          \ {\rm e}^{ i{\sf p} \cdot {\sf x}} \phi ({\sf p}).
\end{equation}
Integrating out $h(\sf x)$, we obtain an effective action for slow
modes $\phi_{\Lambda'}$ :
\begin{equation}
    Z = \int{ \cal D} \phi_{\Lambda'} \ {\rm e}^{-S'[\phi_{\Lambda'}]}.
\end{equation}
 Here we keep only the first order terms with respect to the coupling
constant $ u$,
then we obtain
\begin{equation}
    u({\Lambda'}) = u({\Lambda}) \ \exp({-G_s(0)/{K_{\sigma}}^2})\ 
(\frac{\Lambda'}{\Lambda})
^{-2},
\end{equation}
where
\begin{eqnarray}
  G_{c,s}(\sf x)&=& \langle h_{c,s}^{\pm}({\sf x}) 
h_{c,s}^{\pm}(0)\rangle_h \nonumber \\
      &=& \int{ \cal D}h\    h_{c,s}^{\pm}({\sf x}) h_{c,s}^{\pm}(0)\ 
{\rm e}^{-S[h]}\  /\
    \int{ \cal D}h \ {\rm e}^{-S[h]}    \nonumber \\
      &=& K_{\rho,\sigma} \frac{|d{\Lambda}|}{\Lambda} J_0({\Lambda}|\sf x|),
\end{eqnarray}
and $|d{\Lambda}|={\Lambda}-{\Lambda'}$.

So RG equation is written as
\begin{equation}
   \frac{{\rm d}u}{{\rm d}l} = (2- \frac{1}{K_{\sigma}})\ u,
\end{equation}
  where  ${\rm e}^l= \Lambda/\Lambda'$.
  It shows that when $K_{\sigma}<1/2$, the tunneling is irrelevant;
 and when $K_{\sigma}>1/2$,
it is relevant. In the following we consider both cases.

 First we consider the case of $\nu(=1/q)<1/2$ independent Laughlin state.
Since the irrelevancy of the tunneling, the layer which is applied
source-drain voltage is described by the fixed point which corresponds
to the single layer QH systems
in the long wave length and low energy limit.
Then in this limit the Hall conductance and two-terminal conductance 
are given by $G_{\rm H} = G_{\rm T} = \nu  e^{2}/h$.
At finite temperature there should be corrections
 to two-terminal conductance.

This correction can be obtained by calculating the tunneling current
between the layers by Fermi's golden rule:
\begin{eqnarray}
  I_{{\rm tun}} & =& 2\pi e  \sum_n |\langle n| H_{{\rm tun}} 
|o\rangle|^2
\delta(E_n-E_o-eV)
 \nonumber  \\
       & =& \frac{t^2e }{2\pi} \int {\rm d}x {\rm d}x' {\rm d}{\omega}
                   \   G_{\uparrow}^>(x-x',\omega -eV)  \nonumber \\
        && \ \ \ \ \ \ \ \ \ \ \ \ \ \ \ \ \ \ \ \ \
times G_{\downarrow}^<(x-x',\omega),
\end{eqnarray}
 where $G^<_I$ and $G^>_I$ are the Green's functions.
  These can be expressed as
\begin{eqnarray}
  G^>_I(x,\omega) &=& \int {\rm d}t \ {\rm e}^{{\rm i}\omega t}\
  \langle\psi^{\dagger}_I(x,t) \psi_I(0,0)
\rangle.
\end{eqnarray}
 This is related to the imaginary-time Green's
function
\begin{equation}
 G_I(x, \tau) = \langle T_{\tau}[\psi^{\dagger}_I(x,\tau) \psi_I(0,0)] 
\rangle
\end{equation}
  via analytic continuation, $ G^>_I(x,t)=G_I(x,\tau \to {\rm i}t)$.

In terms of conformal theory, electron operator is the vertex operator
which has conformal dimension $\Delta =1/2\nu$.
 So we have
\begin{equation}
  \langle \psi^{\dagger}_I(z)\psi_I(z') \rangle \ 
 \sim \ \frac{1}{(z-z')^{1/\nu}}
\end{equation}
  where $z=x-{\rm i}c\tau$, and $c$ is velocity of single layer edge mode.

Therefore tunneling density is given as
\begin{equation}
 G^>_I(x,\omega) \sim {\omega}^{1/{\nu}-1}\ {\rm e}^{{\rm i}{\omega}
(x-x'-{\rm i}\alpha)/c  }
   \theta(\omega),
\end{equation}
and one obtains
\begin{equation}
  I_{{\rm tun}} \sim \int_{0}^{eV} {\rm d}\omega\  \omega^{2/\nu-4}
 \sim V^{2/\nu-4}V.
\end{equation}
This tunneling current gives rise to backscattering of the current.
At the finite temperature $T$, we can put the voltage $V$ to $T$.
So the the correction of two-terminal conductance
is given as $G_{\rm T} - \nu e^{2}/h  \propto-  T^{2q-4}$.
 This result is consistent with our renormalization group analysis.

  In the other cases: $\nu=1>1/2$ independent Laughlin state and standard 
hierarchical state, $u$ scales to a large value, then our
RG equation Eq.(15) cannot be relied on.
However, $u$ should scale to a finite value.
The vicinities of the electrodes are non equiliblium, and Hall conductance
 varies
from $2\nu e^2/h$, when $u$ is scaled to large value, to $\nu e^2/h$. Since 
these behavior at $\nu=1$ is investigated in ref.(4), we do not discuss
 them in the present paper.

    Far from the source and drain electrodes, the systems are expected to 
 achieve the equiliblium state in which chemical potentials of the 
two layers are equal.
 Actually $\phi_s$ is massive.
To see this, we suppose $u$ is infinite.
Because of dominancy of $S_{\rm tun}$, $\phi_s$ is given by a
solution of the equation of motion, and it is constant.
Then spin mode has an infinite gap.
Even when $u$ is finite, spin mode has a finite gap.
  Therefore low-energy excitations are charged modes only.
  This means that it is rather stable to flow current on both
 edges of layers
under the perturbation of the source-drain voltage,
 and it can be concluded
that this situation is an equiliblium state.
Therefore if one first assigns a definite value of the current, then the
measured voltage is half of the single layer case.
So the Hall conductance on this area
at $\nu=1>1/2$ independent Laughlin state and spin singlet state must be 
$ G_{\rm H}=2\nu e^2/h$  according
to the Landauer-Buttiker's picture.
 
 So far we have discussed tunneling due to the uniform overlap of the wave 
functions.
Now we consider the effect of impurity scattering.
First let's consider a situation where impurities are sparse.
Tunneling at each impurities occurs independently. We represent the outcome
of such tunneling by a single effective impurity at the each side of 
the edge.
\cite{r1} 
This situation resemble Grayson's experiment.\cite{grayson}
In this case the tunneling amplitude can be written as point
 contact type \cite{r2}
\begin{equation}
  t(x)^{u,l} = t \ \delta(x-x_0^{u,l}),
\end{equation}
where $x_0^{u,l}$ is the position of the impurity.
 Then the tunneling term $S_{\rm tun}$ is replaced by
\begin{equation}
  S_{{\rm tun}} = u\int{\rm d}\tau \cos\frac{\phi_s^+(0,\tau)}{K_\sigma}
\cos\frac{\phi_s^-(0,\tau)}{K_{\sigma}}
\end{equation}  where we put $x_0^u=x_0^l=0$.
To calculate the scaling dimension, we consider effective 
action $S^{{\rm eff}}$, which is
defined as \cite{kf}
\begin{eqnarray}
  Z &=& \int {\cal D} \phi \  \exp(-S_0 - S_{{\rm tun}}[\phi_s(o,\tau)]) 
\nonumber \\
     &=&  \int {\cal D} \phi {\cal D} \theta {\cal D} \lambda 
  \ \   \exp[-S_0[\phi]
-S_{\rm tun}[\theta(\tau)] \nonumber \\  & &
\ \ \ \ \  + \int {\rm d}\tau \lambda(\tau) 
\left( \theta(\tau)-\phi_s(0,\tau)\right)  ] \nonumber \\
&=& \int {\cal D} \theta \ \exp(-S^{{\rm eff}}[\theta]).
\end{eqnarray}
After integrating out, we obtain
\begin{eqnarray}
 S^{{\rm eff}}[\theta] &=& \sum_{\omega} \frac{{\omega}}{2\pi K_{\sigma}}
\left(
 \theta^{+}(-\omega) \theta^{+}(\omega) + \theta^{-}(-\omega) 
\theta^{-}(\omega)
 \right)  \nonumber  \\
  & & \ \ \ \ \ \ \ \ \ \ \ \ +\  u \int {\rm d} \tau\ 
 \cos( \frac{{\theta}^+}{K_{\sigma}} )
 \cos( \frac{{\theta}^-}{K_{\sigma}} ).
\end{eqnarray} 
 As before we integrate out the fast mode to 
obtain 
\begin{equation}
  u(\Lambda') = u(\Lambda)\  
\exp({-\frac{1}{K_{\sigma}^2}G(0)})\ (\frac{\Lambda'}{\Lambda})^{-1},  
\end{equation}
 where
\begin{equation} 
G(\tau) = K_{\sigma}\frac{|d\Lambda|}{\Lambda}{\rm e}^{{\rm i}\Lambda \tau}.
\end{equation}
 Then the RG equation reads
\begin{equation}
  \frac{{\rm }du}{{\rm d}l} = (1-\frac{1}{K_{\sigma}})u.
\end{equation}
Thus if $\nu <1$ tunneling is irrelevant for independent Laughlin 
state and two-terminal conductance behaves as
$
    G_{\rm T}(T) = \nu e^{2}/h - \alpha T^{2/\nu - 2}
$ similarly to the point contact experiment.\cite{mill}

 For the other cases, tunneling is marginal.
 When $\nu=1$, interaction can be neglected, so we can use the 
Landauer-Buttiker transport
theory. Then the electrons partially transmit between the two layers,
and Hall conductance is almost independent of temperature. 

Next we consider a situation where impurities are not sparse, and tunneling 
at each impurities can not be treated independently.
 This case is modeled by
random Gaussian distribution for $t(x)$ with
\begin{equation}
  P_{[t(x)]}  = \exp\left( -\frac{1}{2W} \int {\rm d}x \  t(x)^2 \right).
\end{equation}
Then
\begin{equation}
      \overline{t(x)t(x')}   = W \delta(x-x').
\end{equation}
 The effect of this potential is investigated by Gimarchi and Schulz 
in the theory of
localization of interacting one dimensional electrons systems,\cite{gs} 
and by Kane, Fisher, and Polchinski in QH edge.
\cite{kfp}
In order to treat this potential, we use the reprica trick,\cite{edw}
Noting the relation;
\begin{equation}
  \ln Z = \lim_{n\rightarrow 0} \frac{1}{n}(Z^n-1),
\end{equation}
 when  we consider the average of free energy,
we calculate $\overline{ Z^n} $ in place of $\overline{\ln Z} $
, where $n$ is the number of repricas.
The former average is written as
\begin{equation}
  \overline{ Z^n} = \int (\prod_i {\cal D}\phi_i)\ 
\overline{\exp(-\Sigma_i S[\phi_i])}.
\end{equation}
Then effective action can be expressed as
\begin{equation}
\overline{\exp(-\Sigma_i S [\phi_i])} = \exp(-S^{av}),
\end{equation}
\begin{eqnarray}
   S^{av}=   \sum_i S_0[\phi_i] \ \ \ &&   \nonumber  \\
   -  \sum_{ij}  2W \int {\rm d}\tau&& {\rm d}\tau' {\rm d}x \ 
\cos (\frac{\phi_{s,i}^+(x,\tau) 
- \phi_{s,j}^+(x,\tau')}{K_{\sigma}})  \nonumber \\
& &\times  
\cos(\frac{\phi_{s,i}^-(x,\tau)-\phi_{s,j}^-(x,\tau')}{K_{\sigma}}).
\end{eqnarray}
In this case $W$ plays a role of a coupling constant. According to 
refs.(10) and (19)
RG equation of $W$ is written as
\begin{equation}
 \frac{{\rm d}W}{{\rm d}l} =  ( 3 - \frac{2}{K_{\sigma}}) W.
\end{equation}
 If $K_{\sigma} < 2/3$ it is irrelevant,
then the values of Hall conductance does not change at zero temperature.
On the other hand, if $K_{\sigma} =1 > 2/3$ randomness is relevant. 
As  mentioned above $\phi_s$ is 
massive, then Hall conductance is $ G_{\rm H} = 2 e^2/h $ 
at zero temperature.

 In this paper we have shown the following.

(a) In the case of independent Laughlin state at $\nu<1/2$ (1/3,1/5,...),
 the Hall conductance at zero
    temperature is equal to that of the single layer under 
any mechanism of tunneling.
   At low temperature
    $T$-dependence of the two-terminal conductance is given as
\begin{equation}
    G_{\rm T}(T) = \nu \frac{e^2}{h} - \alpha T^{2/\nu -4},
\end{equation}
for the ideal uniform tunneling case,
\begin{equation}
     G_{\rm T}(T) = \nu \frac{e^2}{h} - \alpha T^{2/\nu -2},
\end{equation}  for the local impurities, and
\begin{equation}
       G_{\rm T}(T) = \nu \frac{e^2}{h} - \alpha T^{4/\nu -6},
\end{equation} for the random impurities,
where $\alpha$ depends on the details of the sample.

(b) In the other case: $\nu=1$ and standard hierarchical state at any $\nu$.
In the presence of the uniform overlap of the wave functions or random 
impurities, Hall conductance on the middle region of the sample is given as
\begin{equation}
      G_{\rm H} = 2 \nu \frac{e^2}{h},
\end{equation}  
because of pseudo-spin gap.
 If the local impurities cause tunneling, two-terminal conductance 
is non universal and almost independent of temprature, since tnnneling is 
marginal.

 All above arguments apply only to $\nu=1/$(odd-integer) QH state.  
  We will study the transport phenomena in hierarchical
\cite{had}
%
and
 paired \cite{had2,mr}  QH states, and finite
size correction is to be discussed later.
\vspace{10mm}

\section*{Acknowledgement}

  We are grateful to K. Imura, Y. Morita, and M. Ogata for useful discussion.
This work is supported by Grant-in-Aid for Scientific Research (C) 10640301
from the Ministry of Education, Science, Sports and Culture.

\end{multicols}

\begin{thebibliography}{99}
\bibitem{note1} In the actual system $v_c$ should be larger than $v_s$ due
to the long-range Coulomb interaction.  The value of these velocities will
not affect our discussion here.
\bibitem{r1} If the local impurity exists on only one side, the transport 
propertys are not effected.
\bibitem{r2} We regard the intensity of the tunneling amplitude as same on the 
both edges, because the difference of them does not make any influence to the
transport.
\end{thebibliography}
\end{document}